\documentclass[preprint,12pt]{elsarticle}

\usepackage{psfrag}
\usepackage{multirow}
\usepackage{pstricks}
\usepackage{color}

\usepackage{amsmath}
\usepackage{epsfig}
\usepackage{amsfonts}
\usepackage{amssymb}

\journal{Physics Letters B}

\begin{document}

\begin{frontmatter}

\title{An effective approach to same sign top pair production at the LHC \\
and the forward-backward asymmetry at the Tevatron}

\author{C. Degrande$^a$, J.-M. G\'erard$^a$, C. Grojean$^{b,c}$, F. Maltoni$^a$, G. Servant$^{b,c}$}

\address{$^a$Centre for Cosmology, Particle Physics and Phenomenology (CP3)\\
Chemin du Cyclotron 2, Universit\'e Catholique de Louvain, Belgium,\\
$^b$CERN Physics Department, Theory Division, CH-1211 
Geneva 23, Switzerland\\
$^c$Institut de Physique Th\'eorique, CEA/Saclay, F-91191 
Gif-sur-Yvette C\'edex, France
}

\begin{abstract}
We study the phenomenology of same sign top pair production at the LHC in a model-independent way. The complete set of dimension six operators involving two top (or anti-top) quarks is introduced and the connection with all possible $t$- or $s$-channel  heavy particle exchanges is established. Only in the former case, same and opposite sign top pair production can be related. We find that while current Tevatron data disfavor $t$-channel models, other production mechanisms are viable and can be tested at the LHC. 
\end{abstract}

\begin{keyword}

Collider physics, top quark, effective approach

\end{keyword}

\end{frontmatter}

\section{Introduction}

The top quark is the only known fermion whose mass resides at the electroweak scale. For this and several other very peculiar phenomenological properties, such as its marginal sensitivity to low scale QCD effects, it is always considered special from both the theoretical  and experimental point of views~\cite{Bernreuther:2008ju,Han:2008xb}.  In many extensions of the Standard Model (SM) the top quark plays a unique role just because it strongly interacts in the fermion mass generation sector and, as a result, quantum effects on precision observables can be large. In this case  new physics must  exist at roughly the same scales that stabilizes such strong effects. It might be related to new degrees of freedom coupling to the top quark. From the experimental point of view, very distinctive signatures at hadron colliders can be exploited to look for new particles associated to its production or decay. This is of course natural if the masses of such new hypothetical states are in the range accessible at the present collider energies, {\it i.e.}, at the Tevatron and LHC. However, it is important to consider also the possibility that these states are slightly heavier and cannot be produced on shell.  In this case new degrees of freedom enter only at the virtual level to modify the production and/or decay properties of the top quark. One well known candidate for such effects is the forward-backward asymmetry (A$_{FB}$) measurement at the Tevatron~\cite{Aaltonen:2011kc}, possibly the first hint for new physics in the top sector. This is exactly where a model-independent approach based on effective operator description is mostly useful~\cite{Jung:2009pi,Zhang:2010dr,AguilarSaavedra:2010zi,Degrande:2010kt,Blum:2011up,Delaunay:2011gv,AguilarSaavedra:2011vw}.

Recently, all the operators that could describe new physics effects in top pair production both at the Tevatron and the  
LHC have been classified~\cite{Kumar:2009vs,Zhang:2010dr,AguilarSaavedra:2010zi,Degrande:2010kt}, 
including those affecting A$_{FB}$. At the LHC, the low probability to have a quark-antiquark initial state  
prevents large contributions from the four-fermion operators. In this work we start from the known observation  that this issue is avoided for same sign top pair production and we perform a complete analysis, including the possible relation with resonant states.

After introducing a complete basis of effective operators, we compute the total cross-section, the invariant mass distribution $m_{tt}$ and the spin  correlations in a model independent way. All possible $t$- and $s$-channel exchanges of heavy particles are expressed in terms of those dimension-six operators. The former may link same and opposite sign top pair productions. However, as we will show, it is disfavored when  the cross-section \cite{Cerrito:2010sv}, the A$_{FB}$~\cite{Aaltonen:2011kc}  and the invariant mass distribution \cite{Goldschmidt:2010} measurements are considered simultaneously.

\section{The operators}

Any operator contributing to same sign top pair production can be expressed as a linear combination of 
\begin{eqnarray}
 \mathcal{O}_{RR} &=& \left[\bar{t}_R\gamma^\mu u_R\right]\left[\bar{t}_R\gamma_\mu u_R\right]\nonumber\\
 \mathcal{O}_{LL}^{(1)} &=& \left[\bar{Q}_L\gamma^\mu q_L\right]\left[\bar{Q}_L\gamma_\mu q_L\right]\nonumber\\
 \mathcal{O}_{LL}^{(3)} &=& \left[\bar{Q}_L\gamma^\mu \sigma^a q_L\right]\left[\bar{Q}_L\gamma_\mu \sigma^a q_L\right]\nonumber\\
 \mathcal{O}_{LR}^{(1)} &=& \left[\bar{Q}_L\gamma^\mu q_L\right]\left[\bar{t}_R\gamma_\mu\ u_R\right]\nonumber\\
 \mathcal{O}_{LR}^{(8)} &=& \left[\bar{Q}_L\gamma^\mu T^A q_L\right]\left[\bar{t}_R\gamma_\mu\ T^A u_R\right]\label{op}
\end{eqnarray}
with the three Pauli matrices normalized as ${\rm Tr}\left(\sigma^a\sigma^b\right)=2\delta^{ab}$ and the eight $SU(3)_c$ generators normalized as ${\rm Tr}\left(T^AT^B\right)=\frac{1}{2}\delta^{AB}$. In Eq.~\eqref{op}, $Q_L$ ($q_L$) is the left-handed doublet of the third (first) quark generation. The relevant effective Lagrangian is then given by 
\begin{eqnarray}
 \mathcal{L}_{\rm dim=6}^{qq\to tt}&=& \frac{1}{\Lambda^2}\left(c_{RR}\mathcal{O}_{RR} +c_{LL}^{(1)}\mathcal{O}_{LL}^{(1)} + c_{LL}^{(3)}\mathcal{O}_{LL}^{(3)}\right.\nonumber\\
&&\left. + c_{LR}^{(1)}\mathcal{O}_{LR}^{(1)} + c_{LR}^{(8)}\mathcal{O}_{LR}^{(8)}\right)+h.c..
\label{eq:ltt}
\end{eqnarray}
The $\mathcal{O}_{LL}^{(1)}$ and $\mathcal{O}_{LL}^{(3)}$ operators contain the same product of neutral currents $\left[\bar{t}_L\gamma^\mu u_L\right]\left[\bar{t}_L\gamma_\mu u_L\right]$, which are relevant for $uu\to tt$.  In addition they  contain $\left[\bar{b}_L\gamma^\mu d_L\right]\left[\bar{b}_L\gamma_\mu d_L\right]$, which can  contribute to the $B_d$ mixing and to di-jet production. For example,  the linear combination $c_{LL} = c_{LL}^{(1)}+c_{LL}^{(3)}$ can be constrained  from the former \cite{Bona:2007vi}  
\begin{equation}
 \left|c_{LL}\right|\left(\frac{1\, \text{TeV}}{\Lambda}\right)^2<2.1\times 10^{-4}.\label{UTfit}
\end{equation}
The difference between the two $LL$ operators in Eq.~\eqref{op} is thus in the product of charged currents $\left[\bar{t}_L\gamma^\mu d_L\right]\left[\bar{b}_L\gamma_\mu u_L\right]$ present only in $\mathcal{O}_{LL}^{(3)}$ and affecting the top decay as well as single top production \cite{Zhang:2010dr}.

\section{Phenomenology}

\begin{figure}[!ht]
  \centering
\includegraphics[width=0.25\textwidth]{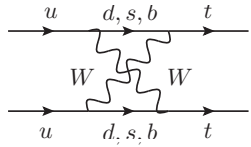}
\caption{SM contribution to $uu\to tt$}
\label{fig:SM}
\end{figure}

\begin{figure}[!htb]
  \centering
    \includegraphics[width=0.45\textwidth]{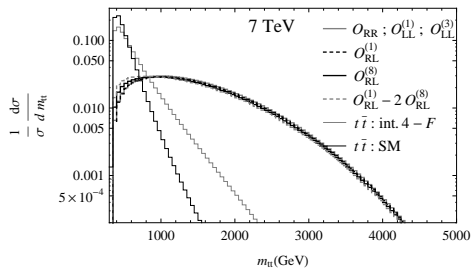}
    \caption{Normalized invariant mass distribution for same sign top pair production at the LHC. The distribution can be trusted only for $m_{tt}\ll\Lambda$. The interference between the SM and the four-fermion operators as well as the SM for $t\bar{t}$ production are also displayed for comparison.}
  \label{fig:mtt}
\end{figure}

\begin{figure}[!htb]
  \centering
    \includegraphics[width=0.45\textwidth]{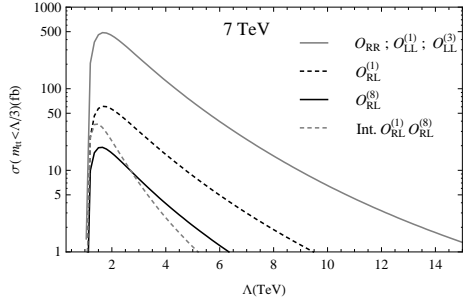}
    \caption{Cross-section of $pp\to tt$ at the LHC with an upper cut on the invariant mass at $\frac{\Lambda}{3}$ for $c_i=1$. Parameters: $m_t=174.3$ GeV, CTEQ6L1 pdf set~\cite{Nadolsky:2008zw}, $\mu_F=\mu_R=m_t$.}
  \label{fig:cs}
\end{figure}

At the partonic level, the leading order cross-section for same sign top pair production is given by
\begin{eqnarray}
 \frac{d\sigma}{dt} &=& \frac{1}{\Lambda^4}\left[\left(\left|c_{RR}\right|^2+\left|c_{LL}\right|^2\right)\frac{\left(s-2m_t^2\right)}{3\pi s}\right.\nonumber\\
&&+ \left(\left|c_{LR}^{(1)}\right|^2+\frac{2}{9}\left| c_{LR}^{(8)}\right|^2\right) \frac{\left(m_t^2-t\right)^2+\left(m_t^2-u\right)^2}{16 \pi s^2}\nonumber\\
&&-\left.\left(\left|c_{LR}^{(1)}\right|^2+\frac{8}{3}\Re\left(c_{LR}^{(1)}{c_{LR}^{(8)}}^*\right)-\frac{2}{9}\left| c_{LR}^{(8)}\right|^2\right)\frac{m_t^2}{24 \pi s}\right].\nonumber\\
\label{xsection}
\end{eqnarray}
The dominant contribution to this cross-section is due to the new physics amplitudes squared because the one-loop SM process depicted in Fig.~\ref{fig:SM} is strongly suppressed by the squares of the $V_{ub}$ CKM matrix element and of the bottom quark mass. Lowest order contributions are thus $\mathcal{O}\left(\Lambda^{-4}\right)$ contrary to opposite sign top pair production for which the largest corrections arise from the $\mathcal{O}\left(\Lambda^{-2}\right)$ interference. After integration over $t$, the cross-section grows like $s$ as expected from dimensional analysis. In fact, only the interference between the $LR$ operators is proportional to $m_t^2$, see Eq.~\eqref{xsection}, and does not have this behaviour. As a  consequence, a large part of the total cross-section at the LHC comes from the region where $m_{tt}\sim1$ TeV as shown on Fig.~\ref{fig:mtt}. In this region, however, the $1/\Lambda$ expansion cannot be trusted for
values of $\Lambda$  around 1~TeV we consider in our study. There is no such concern at the Tevatron as the $m_{tt}$ distribution is peaked instead below 500 GeV. 
Figure~\ref{fig:cs} displays the cross-section with a upper cut on $m_{tt}$ at $\Lambda/3$ as a function of $\Lambda$ for $c_i=1$, where $c_i$ is a generic label for the coefficients in Eq.~\eqref{eq:ltt}. This choice ensures that the $m_{tt}$ distribution is at most about 20\% below (above) its true value for an $s$- ($t$-) channel exchange. The general case can be easily inferred since the coefficient dependences factorise in Eq.~\eqref{xsection}. At 14 TeV, the cross-section increases by a factor 2 for $\Lambda\sim2$ TeV up to a factor 4 for $\Lambda\sim14$ TeV.


Figure~\ref{fig:mtt} shows that the $m_{tt}$ shapes given by the different operators, appear to be quite similar. The maximal effect of the interference term corresponds approximatively to the linear combination $\mathcal{O}_{LR}^{(1)} -2\mathcal{O}_{LR}^{(8)}$. As foreseen, the interference can only give a sizeable effect for low $m_{tt}$ since it does not grow with $s$. Again, there are no significant changes at 14 TeV. The distribution is only stretched to the higher invariant mass region.

In contrast with the $m_{tt}$ distribution, the spin correlations provide in principle a very efficient  observable to discriminate among the contributions from the various operators in Eq.~\eqref{op}. The main reason is that the latter have a well defined chirality structure and no interference with the Standard Model is possible. Let us define the normalized differential $tt$ cross-section
\begin{eqnarray}
\frac{1}{\sigma}\frac{d\sigma}{d\cos\theta_1d\cos\theta_2} &=& \frac{1}{4}\left[1+C \cos\theta_1 \cos\theta_2\right.\nonumber\\ &&\left.+b\left(\cos\theta_1 +\cos\theta_2\right)\right]\,,
\end{eqnarray}
where $\theta_1$ $(\theta_2)$ is the angle between the momentum in the top rest frame of the charged lepton resulting from the first (second) top decay and the top momentum in the $tt$ rest frame. Then, the $C$ and $b$ parameters can be directly computed from the helicity cross-sections, namely
\begin{eqnarray}
C&=&\frac{1}{\sigma}\left(\sigma_{++}+\sigma_{--}-\sigma_{+-}-\sigma_{-+}\right)\nonumber\\
b&=&\frac{1}{\sigma}\left(\sigma_{++}-\sigma_{--}\right)\,,
\end{eqnarray}
where the first (second) indice refers to the helicity of the first (second) top quark. For $\mathcal{O}_{RR}$, $C=1$ and $b=0.997$. For $\mathcal{O}_{LL}^{(1)}$ and $\mathcal{O}_{LL}^{(3)}$, only the sign of $b$ changes. The two remaining operators in Eq.~\eqref{op} are caracterized by $C\approx 1$ and $b\approx 0$. $C$ and $b$ are here calculated on the full cross-section, {\it i.e.}, without any cut on $m_{tt}$. However, $C=1$ for $\mathcal{O}_{RR}$, $\mathcal{O}_{LL}^{(1)}$ and $\mathcal{O}_{LL}^{(3)}$ is independent of such a cut. As a result, such strong spin correlations could be used to enhance the sensitivity to the signal and to identify the possible contributing operators.

\section{Link with a t-channel exchange}

Flavor changing $t$-channel exchanges invoked to account for the Tevatron  A$_{FB}$ might imply a large same sign top pair production at the LHC \cite{Berger:2011ua,Cao:2011ew}. Table~\ref{tab:tchannel} shows the coefficients of the operators of Eq.~\eqref{op} for any possible particle exchange in the $t$-channel. If the new physics is in the reach of the LHC, the left coupling of all vectors have to be very tiny to satisfy Eq.~\eqref{UTfit}. 

\renewcommand{\arraystretch}{1.2}
\begin{table}[h]
\begin{center}
\begin{tabular}{|c|c|c|c||c|c|c|c|c|}
\hline
Spin&SU(3)&SU(2)&Y&$c_{RR}$&$c_{LL}^{(1)}$&$c_{LL}^{(3)}$&$c_{LR}^{(1)}$&$c_{LR}^{(8)}$\\
\hline
1&1&1&0&$-\frac{1}{2}$&$-\frac{\xi^2}{2}$&&$-\xi$&\\
1&8&1&0&$-\frac{1}{6}$&$-\frac{\xi^2}{24}$&$-\frac{\xi^2}{8}$&&$-\xi$\\
0&1&2&$\frac{1}{2}$&&&&$-\frac{1}{6}\xi$&$-\xi$\\
0&8&2&$\frac{1}{2}$&&&&$-\frac{2}{9}\xi$&$\frac{1}{6}\xi$\\
1&1&3&0&&&$-\frac{\xi^2}{2}$&&\\
1&8&3&0&&$-\frac{3}{8}\xi^2$&$\frac{5}{24}\xi^2$&&\\
\hline
\end{tabular}
\end{center}
\caption{Coefficients of the operators up to an overall factor $g_R^2$ for all possible $t$-channel  particle exchanges (of mass $M=\Lambda$) identified by their quantum numbers ($Q=T_3+Y$). $\xi=\frac{g_L}{g_R}$ with $g_L$ ($g_R$) the coupling to the density $\bar{t}_Rq_L$ ($\bar{Q}_Lu_R$) or to the current $Q_L\gamma^\mu q_L$ ($t_R\gamma^\mu u_R$). The normalization for the $SU(2)_L$ and $SU(3)_c$ currents are the same as in Eq.~\eqref{op}.} 
\label{tab:tchannel}
\end{table}

\begin{figure}[h]
  \centering
\includegraphics[width=0.27\textwidth]{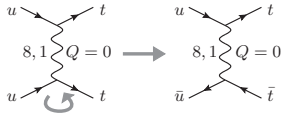}
\caption{Possible connection between same and opposite sign top pair production through a $t$-channel self-conjugate particle exchange.}
\label{fig:tchan}
\end{figure}

In general, no  relation exists between same and opposite sign top pair production. Consequently, the five different parameters present in \eqref{xsection} cannot be related to those affecting the total cross-section, $m_{t{\bar t}}$ ($c_{Vv}$ and $c_{Vv}'$) and A$_{FB}$ ($c_{Aa}$ and $c_{Aa}'$) in $q\bar{q}\to t\bar{t}$ ,{\it i.e.}, \cite{Degrande:2010kt}
\begin{eqnarray}
\mathcal{L}^{q\bar{q}\to t\bar{t}}&=&\left(\frac{c_{Vv}}{2}\pm \frac{c_{Vv}'}{4}\right)\left[\bar{t}\gamma_\mu T^at\right]\left[\bar{q}\gamma_\mu T^aq\right]\nonumber\\
&+&\left(\frac{c_{Aa}}{2}\pm \frac{c_{Aa}'}{4}\right)\left[\bar{t}\gamma_\mu\gamma_5 T^at\right]\left[\bar{q}\gamma_\mu\gamma_5 T^aq\right]\,,\label{eq:ttbar}
\end{eqnarray} 
where the upper (lower) sign should be taken for the up (down) quark in $q$. Yet in the special case of a flavor changing $t$-channel, each vertex can be replaced by its hermitian conjugate (see Fig.~\ref{fig:tchan}) if the exchanged particle is self-conjugate. The expression of the coefficients relevant for $t\bar{t}$ are displayed in Tab.~\ref{tab:ttbar} for the allowed cases. 
\begin{table}[h]
\begin{center}
\begin{tabular}{|c|c|c||c|c|c|c|}
\hline
Spin&SU(2)&Y&$c_{Vv}$&$c'_{Vv}$&$c_{Aa}$&$c'_{Aa}$\\
\hline
1&1&0&$-\frac{1}{2}$&$-1$&$-\frac{1}{2}$&$-1$\\
0&2&$\frac{1}{2}$&$-\frac{1}{2}\left(\left|\xi\right|^2+\frac{1}{2}\right)$&$-\frac{1}{2}$&$\frac{1}{2}\left(\left|\xi\right|^2+\frac{1}{2}\right)$&$\frac{1}{2}$\\
\hline
\end{tabular}
\end{center}
\caption{Expression of the parameters relevant for $t\bar{t}$ up to an overall factor $\left|g_R\right|^2$ for a color singlet particle of mass $M=\Lambda$ in the $t$-channel. The coefficients for the corresponding color octets are obtained by multiplying them all by $-\frac{1}{6}$.} 
\label{tab:ttbar}
\end{table}
\begin{figure*}[!ht]
  \centering
    \includegraphics[width=0.3\textwidth]{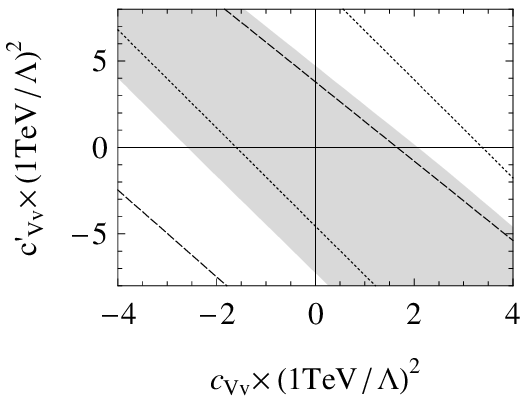}
    \includegraphics[width=0.3\textwidth]{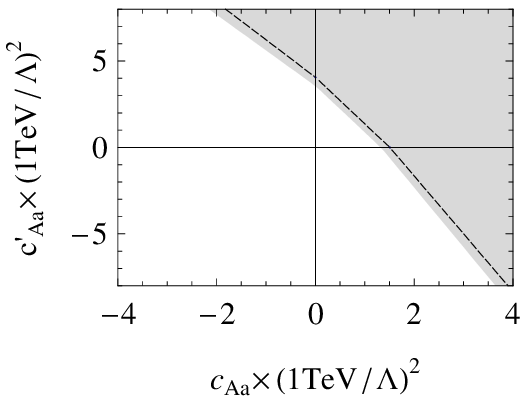}
    \includegraphics[width=0.3\textwidth]{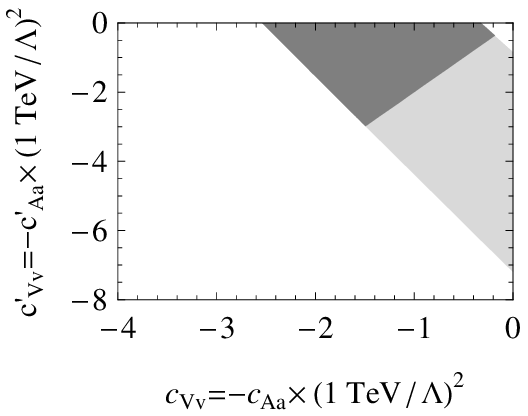}
    \caption{On the left, in gray, the region for $c_{Vv}$ and $c'_{Vv}$ allowed at 95\% by the cross-section (delimited by dotted line) and the shape of the invariant mass distribution (delimited by dashed line). In the center, in gray, the region for $c_{Aa}$ and $c'_{Aa}$ allowed at 95\% by A$_{FB}$ for $m_{t\bar{t}}<450$ GeV (the full region plotted is allowed) and for $m_{t\bar{t}}\ge 450$ GeV (above the dashed line) \cite{Aaltonen:2011kc}. On the right, the allowed region by all these observables for  $c_{Vv}=-c_{Aa}$ and $c'_{Vv}=-c'_{Aa}$ which corresponds to the still allowed spin 0 case (see Table~\ref{tab:ttbar}). The dark gray region can only be obtained for a $t$-channel scalar.}
  \label{fig:exclusion}
\end{figure*}

The $t$-channel models are already disfavored by the Tevatron data due to the relation between the vector and axial coefficients ($\left|c_{Vv}\right|=\left|c_{Aa}\right|$ and $\left|c_{Vv}'\right|=\left|c_{Aa}'\right|$). On the one hand, the agreement of the measured total cross-section and the $m_{t{\bar t}}$ distribution with the SM predictions requires $c^{(\prime)}_{Vv}$ to be small as shown on  Fig.~\ref{fig:exclusion}. On the other hand, the observed deviation for A$_{FB}$ \cite{Aaltonen:2011kc} implies that $c^{(\prime)}_{Aa}$ should be large. In fact, the color singlet vector \cite{Jung:2009jz} and the color octet scalar are immediately ruled out since they give the wrong sign for A$_{FB}$~\cite{Degrande:2010kt},
\begin{eqnarray}
&&\delta A\left(m_{t{\bar t}}<450\,\text{GeV}\right)=\qquad\qquad\qquad\qquad\qquad\qquad\qquad\nonumber\\
&&\qquad\qquad\left(0.023^{+3}_{-1} c_{Aa} +0.0081^{+6}_{-4} c'_{Aa}\right) \left(\frac{1\,\text{TeV}}{\Lambda}\right)^2\nonumber\\
&&\delta A\left(m_{t{\bar t}}\ge450\,\text{GeV}\right)=\qquad\qquad\qquad\qquad\qquad\qquad\qquad\nonumber\\
&&\qquad\qquad\left(0.087^{+10}_{-9} c_{Aa} +0.032^{+4}_{-3} c'_{Aa}\right) \left(\frac{1\,\text{TeV}}{\Lambda}\right)^2.
\end{eqnarray}
After combining all the constraints, we conclude that a color octet vector is also excluded while a small region, depicted in Fig.~\ref{fig:exclusion}, remains for the case of a color singlet scalar. This region disappears if we change the C.L. to 85\%. All the allowed regions have been obtained similarly as in Ref.~\cite{Degrande:2010kt}. This last case is also constrained for low masses by the Tevatron search for  $tt$ production \cite{Aaltonen:2008hx}. Assuming the same acceptance (0.5\%) we find that $c_i\sim1$ are still allowed.  A very recent analysis based on operators in Eq.~\eqref{op} gives similar constraints \cite{cdf:2011ss}.

We note that when the interference between the new physics and the SM is negative, the new physics squared (NP$^2$) can cancel the effect of the interference on the total cross-section for large value of the coupling or for small mass. It has been shown \cite{Berger:2011ua,Cao:2011ew,Jung:2009jz} that the asymmetry can be explained with a rather light color singlet vector coupled only to the right-handed $u$ and $t$ quarks. Of course, this region of the parameter space cannot be probed in our effective approach. However, the invariant mass distribution shape for a light state in the $t$-channel is also only marginally consistent with the data (Ref.~\cite{Jung:2011zv} suggests, though,  that this problem could be alleviated thanks to a reduced acceptance rate of the top quarks in the forward region). For a color singlet scalar, the NP$^2$ contribution to the asymmetry is negative and  implies that $\delta A\left(m_{t{\bar t}}\ge450\, \text{GeV}\right)\lesssim 0.2$.

\section{Link with an s-channel exchange}

The effects of any heavy qq-resonance (listed in Ref.~\cite{Han:2010rf}) can be approximated by the four-fermion operators \eqref{op} at low energy (see Tab.~\ref{tab:schannel}). 
\begin{figure}[htb]
  \centering
\includegraphics[width=0.35\textwidth]{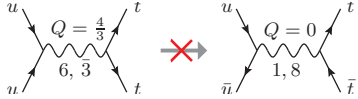}
\caption{Diagrams for same and opposite sign top pair production through an $s$-channel particle exchange.}
\label{fig:schan}
\end{figure}
A color anti-triplet scalar cannot contribute because its coupling is asymmetric under the exchange of the two fermions. It should also be noted that only axial (vector) couplings contribute to the $uu\to tt$ for the color sextet (anti-triplet) iso-doublet resonances. The cases of scalar and vector sextet have been treated in Refs.~\cite{Berger:2010fy,Zhang:2010kr}. In general same sign top pair production through an $s$-channel particle exchange cannot be related to opposite sign top pair production because of color and electric charges (see Fig.~\ref{fig:schan}). For the same reason, opposite sign top pair production through a $u$-channel particle exchange \cite{Shu:2009xf,Arhrib:2009hu,Dorsner:2009mq,Ligeti:2011vt} cannot be related to same sign top pair production. 
\begin{table}[h]
\begin{center}
\begin{tabular}{|c|c|c|c||c|c|c|c|c|}
\hline
Spin&SU(3)&SU(2)&Y&$c_{RR}$&$c_{LL}^{(1)}$&$c_{LL}^{(3)}$&$c_{LR}^{(1)}$&$c_{LR}^{(8)}$\\
\hline
1&$\bar{3}$&2&$\frac{5}{6}$&&&&$-\frac{1}{6}$&$\frac{1}{2}$\\
1&6&2&$\frac{5}{6}$&&&&$-\frac{1}{3}$&$-\frac{1}{2}$\\
0&6&1&$\frac{4}{3}$&$\frac{1}{4}$&&&&\\
0&6&3&$\frac{1}{3}$&&$-\frac{3}{8}$&$-\frac{1}{8}$&&\\
\hline
\end{tabular}
\end{center}
\caption{Coefficients of the operators up to an overall factor $-g_1 g_3$ for all possible $s$-channel exchanges (of mass $M$ and with a coupling $g_1$ ($g_3$) to the first (third) generation quarks) identified by their quantum numbers.} 
\label{tab:schannel}
\end{table}

\section{Conclusion}

Only five independent effective operators of dimension six contribute to same sign top pair production. Among them two operators are already severely constrained by flavour data and cannot play any role in processes  at the TeV scales. The cross sections can be of the order of a pb both at 7 TeV and at 14 TeV if the scale of the new physics is about 2 TeV. LHC searches in the same-sign dilepton channel will be probing these cross sections this year. It makes this channel particularly competitive to search for new physics in the top sector (see also \cite{Rajaraman:2011rw}  for probing like-sign top production using single lepton events). The strong spin correlations can, in principle, be used to distinguish the different operators. We have shown also that the $t$-channel scenarii are disfavored by the Tevatron data. The relation between opposite and same sign top pair productions can then not be directly used to fix the production rate of the latter at the LHC.  On the other hand, the LHC has definitely the potential to constrain the corresponding operators.

{\bf Note added:} 
While this work was being finalized, Ref.~\cite{AguilarSaavedra:2011zy} appeared on the same subject.

{\bf Acknowledgements:} 
The work of C.D., J.-M. G. and F.M. is supported by the Belgian Federal Office for Scientific, Technical and Cultural Affairs through the Interuniversity Attraction Pole No. P6/11. C.D. is a fellow of the Fonds National de la Recherche Scientifique. C.G. is partly supported by the European Commission under the contract ERC advanced grant 226371 MassTeV and the contract PITN-GA-2009-237920 UNILHC. G.S is supported by the ERC Stg Grant Cosmo@LHC.

\bibliographystyle{model1-num-names}
\bibliography{biblio}

\end{document}